\documentstyle [12pt] {report}
\begin{document}
\setlength{\baselineskip}{24pt}

\title{\large \bf  QUANTIZATION OF CLOSED MINI-SUPERSPACE  \\
                                       MODELS AS BOUND STATES}

\author{\bf J. H. Kung \\
Harvard-Smithsonian Center for Astrophysics \\
60 Garden Street \\
Cambridge, MA 02138}

\date{}
\maketitle
\begin{abstract}

Wheeler-DeWitt equation is applied to $k > 0$ Friedmann Robertson Walker
metric with various types of matter.  It is shown that if the Universe
ends in the  matter dominated era (e.g., radiation or pressureless gas)
with zero cosmological constant, then the resulting Wheeler-DeWitt equation
describes a bound state problem.   As  solutions of a non-degenerate
bound state system, the eigen-wave functions are  real (Hartle-Hawking)
and the usual issue associated with the   ambiguity in the boundary
conditions for the wave functions is resolved. Furthermore, as a bound state
problem, there exists a quantization condition that relates the curvature
of the three space with the energy density of the Universe. Incorporating
a cosmological constant in the early Universe (inflation) is given as a
natural explanation for the large quantum number associated with our
Universe, which resulted from the quantization condition.

It is also shown  that if there is a cosmological constant $\Lambda > 0$
in our Universe that persists for all time, then the resulting
Wheeler-DeWitt equation  describes  a non-bound state system,
regardless of the magnitude of the cosmological constant.  As a
consequence, the wave functions are in general complex (Vilenkin) and
the initial conditions for  wave functions are a free parameters not
determined by the formalism.
\end{abstract}

\begin{flushleft}{\large \bf I. INTRODUCTION}
\end{flushleft}
\setcounter{chapter}{1}
\setcounter{equation}{0}

    It is well known that the Hamiltonian formulation of classical
general relativity is a constraint system [1,2].   Mathematically,
the existence of constraints can be understood as a consequence  of the
fact that not all the components of the metric tensor $g_{\mu,\nu}$  are
not dynamical coordinates. In the Arnowitt, Deser, Misner (ADM)
decomposition of the space-time metric [3], the constraints arise
because canonical momenta conjugate to lapse function and shift
vectors vanish identically.  Quantization of  a system via Hamiltonian
goes under the name of  canonical quantization, in which one has the
option of using either  Schrodinger or Heisenberg representation of
operators. Vanishing of the momentum conjugate to the lapse function, in
Schrodinger representation, gives the celebrated Wheeler-DeWitt
equation [4,5]. Here, operators act on ``superspace'' that is  an
infinite-dimensional space of all possible 3-geometries, which are not
related by diffeomorphism,  and all matter field configurations.   In
practice, to make problems analytically tractable, the infinite dimensional
superspace is reduced to a finite dimensional mini-superspace by working in
models with idealized symmetries [6].  In cosmology, the metric of primary
interest is that of Friedmann Robertson Walker (FRW) Universe.  The
gravitational sector  of the corresponding mini-superspace models have only
one degree of freedom, which is  the scale factor of the Universe.

The  operator ordering  problem, which arises in the formalism, has been
addressed in the literature [7-9].   Halliwell [7]  and Misner [8] have
demonstrated that the operator ordering ambiguity in the Wheeler-DeWitt
equation can be completely fixed by demanding invariance under field
redefinition of both the three-metric and the lapse function for $D > 1$,
where $D$ is the number of gravitational coordinates in a mini-superspace
model.  In semiclassical calculations, operator ordering is usually
dictated by convenience.  This is because a wrong factor ordering
introduces  an error that is in higher power of Planck's constant.

A scalar field is usually used in the matter sector of the Wheeler-DeWitt
equation [10-13].  This is because one is particularly interested in
applying the Wheeler-DeWitt equation to elucidate properties of our
early Universe near the Planck epoch, when classical description of
matter and gravity must necessarily break down.

Because the Wheeler-DeWitt equation is a second order differential equation,
a priori the initial conditions (or boundary conditions) are not determined
by the formalism and must be specified to completely describe  a system.
Hartle and Hawking [14,15]  have proposed  ``no boundary'' boundary
condition,  for which the resulting wave functions are real. Hartle-Hawking
condition is essentially a prescription for selecting  four-manifolds
that will be included  in the Euclidean path integral over history. On the
other hand, Vilenkin [16,17]  has proposed a ``tunneling''  boundary
condition, for which the resulting wave functions are complex.  Vilenkin's
condition is a proposal that the wave function should consist solely of
outgoing waves from the initial singularity.

     In this paper, we go against the conventional wisdom  and first study
the mini-superspace model corresponding to closed  FRW metric with zero
cosmological constant  and $\rho= \rho_{{\rm rad}} + \rho_{{\rm dust}}$.
Several comments are in order.    First, it is not physically  inane to
use  a classical description of matter in a quantum mechanical equation.
The situation is analogous to  solving for  atomic wave functions while
using  a classical coulomb interaction. Second, because of the usage of
classical description of matter, the resulting wave functions are
adequate in describing only the  late phase of a universe.  There are
several motivations for studying this particular model.  First, the formalism
for quantum cosmology is still in its rudimentary stage, e.g., the role of
time, interpretation of wave functions, usage of Euclidean continuation,
etc. are not clear [18].   Since the correct interpretation of the
theory should give the classical limit,  we feel that it is  necessary
to look into the consequences of  the  formalism  in this classical regime.
Second, limiting ourselves to  a closed FRW metric is for a technical
reason.  Of the three possible FRW metrics, i.e., closed, flat, and open,
only the closed FRW metric has a finite action.  Third, even though this is
a classical regime, application of  quantum mechanical  formalism can
lead to interesting results.  Indeed,  because  the resulting Wheeler-DeWitt
equation resembles a bound state problem, the  quantization condition gives
an interesting  relation between the energy density of the Universe and the
curvature of the three space. Fourth,  if our Universe  is a closed
universe that has gone through an inflationary phase, in its early epoch
but with zero cosmological constant today, then we may be able to say
something about the nature of a wave function  immediately following the
 inflationary phase by invoking the continuity of the wave function and
its derivative.  Here, we get the conclusion that the appropriate
boundary condition for a wave function is that of Hartle and Hawking.
The  conclusion is based on the fact that  eigen-wave functions for
a nondegenerate bound state system are necessarily  real.

     The layout of the paper is as follows. In Sec. II, some notations
and basic results are quoted from the literature.  In Sec. III,  we solve
Wheeler-DeWitt equation in a closed form  for  the closed FRW metric
with $\rho= \rho_{{\rm rad}}$  to get the complete set of eigen-wave
functions. The analysis is carried out  for the Laplacian and for an
arbitrary operator ordering. In Sec. IV, the analysis is repeated for the
more general case of $\rho = \rho_{{\rm rad}} + \rho_{{\rm dust}}$.
In Sec. V, we study the consequences of including  a cosmological constant
to the Wheeler-DeWitt equation for the  $k>0$  FRW metric with
$\rho = \rho_{{\rm rad}} + \rho_{{\rm dust}}$. In Sec VI, we briefly
discuss generalization of including more gravitational degree of freedom
and its anticipated effects on the nature of wave functions and their
corresponding boundary conditions.

\begin{flushleft}{\large \bf II. PRELIMINARIES}
\end{flushleft}
\setcounter{chapter}{2}
\setcounter{equation}{0}

      The starting point is action for the system,

\begin{equation} I = I_{M} + I_{G}.\end{equation}

\noindent $I_{M}$ is an action for matter, and

\begin{equation}I = -{1\over {16 \pi G}}\int d^4x \sqrt{-g}R\end{equation}

\noindent is the familiar Einstein's action for gravity.   We will be
primarily interested in the quantum mechanics  of FRW metric.

\begin{equation} ds^2 = -dt^2 + a^2\left ( {dr^2\over {1-kr^2}} +
r^2d\Omega_3^2\right ),\end{equation}

\noindent where ($a$) is the scale factor of the Universe, $d\Omega_3^2$ is
the metric on a unit three-sphere, and $k>0, k=0, k<0$ correspond to closed,
flat, and open geometry, respectively.

The ``time-time'' component of  Einstein's equation gives

\begin{equation} {\dot a}^2 + k = {8\pi G\over 3}\rho a^2.\end{equation}

\noindent For FRW metric, the Ricci scalar is

\begin{equation}R = -6\left( {\ddot a\over a} +
\left ( {\dot a\over a}\right)^2 + {k\over a^2} \right).\end{equation}

\noindent For the closed FRW metric,  the integral over the spatial
coordinates in the expression for $I_G$ (2.2) is finite and can be
explicitly done.

\begin{equation}I_G = -{3V_3\over 8\pi Gk^{3/2}}\int dt  a
\left( {\dot a}^2 -k \right),\end{equation}

\noindent where $V_3 = 2\pi^2$ is the volume of a three dimensional space
with  constant positive  ($k=1$) curvature.  We have also integrated by parts.
An  astute reader may question the justification for integrating by parts.
The reasoning is quite involved [2,19].  In turns out that there are two
methods of deriving   Einstein's  equations (e.g. (2.4)) from  the
principle of least action.  In the first method, when the action (2.2)
is  varied,  one needs to require  that  both $\delta \dot g_{\mu\nu} = 0$
and $\delta g_{\mu \nu} = 0 $  at appropriate boundaries in order for all
the resulting  surface terms to vanish.     This is  because, in general,
the Ricci scalar involves terms with $\ddot g_{\mu\nu}$.

      In the second method, one starts with an action that differs from
(2.2)   by a term that  is proportional to the trace of the second
fundamental form at the boundary.  Here, to get Einstein's equation, one
only needs to require  that $\delta g_{\mu \nu}$  vanish on all
appropriate boundaries when the new action is varied.  Mathematically,
the addition of the  new term to the action can be viewed as an alternative
way of enforcing all the  surface terms to vanish in the variation.

In the Hamiltonian formulation of general relativity,  it is the latter
action that is preferred because of the nature of the variational
conditions. For the closed FRW metric, the resulting action for gravity
is precisely (2.6).  So finally, the justification for integrating by
part is that in actuality we are using the latter action.

      From (2.6), the canonical momentum conjugate to  scale factor is

\begin{equation} \pi_a \equiv {\delta I_G\over \delta{\dot a}} =
-{3 V_3\over 4\pi G k^{3/2}}a\dot a = -{3 \pi \over 2 G k^{3/2}}a\dot a .
\end{equation}

\noindent In terms of $\pi_a$, the ``time-time'' component of Einstein's
equation  (2.4) is

\begin{equation}\pi^2_a + \left ({3\pi\over 2 G}\right)^2
{1\over k^3}\left [ k a^2 - {8\pi G\over 3}\rho a^4\right]=0.\end{equation}

\noindent We will not need the expression for the Wheeler-DeWitt equation
in its most general form.  To quantize the system, we replace
$\pi_a \rightarrow -i{\partial  \over \partial a}$ in (2.8) to get

\begin{equation}\left [- a^{-p}{\partial\over \partial a}a^p
{\partial \over \partial a} + \left ({3\pi\over 2 G}\right)^2{1\over k^3}
\left ( k a^2 - {8\pi G\over 3}\rho a^4\right ) \right]\psi (a) = 0.
\end{equation}

\noindent The parameter p represents the ambiguity in the ordering of
$a$ and ${\partial \over \partial a}$.  $p = 1$ is the Laplacian  ordering
[9,20,21].  In the subsequent  use of WKB approximation, a convenient
$p = 2$ will be used.

To describe the  matter sector, we substitute
\begin{equation} {8\pi G\over 3}\rho  \rightarrow {8\pi G\over 3}
\left [ \rho_r ({a_o\over a})^4 + \rho_d({a_o\over a})^3\right] +
{\Lambda\over 3},\end{equation}

 \noindent where $\rho_r$, $\rho_d$, and $\Lambda$ are radiation energy
density, matter (dust) energy density, and cosmological constant, respectively.

It is convenient to define several dimensionless parameters.

\begin{equation}\alpha^2\equiv {3\pi \over 2 G k}\end{equation}

\begin{equation}\gamma \equiv {8\pi G\over 6 k}\rho_d a_o^3\end{equation}

\begin{equation}\beta^2 \equiv ({3\pi\over 2Gk})^2\left( {8\pi G\over 3k}
\rho_r a_o^4 +  ({8\pi G\over 6k}\rho_d a_o^3)^2\right)\end{equation}

\begin{equation} \mu \equiv  ({3\pi\over 2G})^2{\Lambda\over 3k^3}
\end{equation}

\noindent In terms of these variables the Wheeler-DeWitt equation is

\begin{equation} \left [- a^{-p}{\partial\over \partial a}a^p
{\partial \over \partial a} + \alpha^4(a-\gamma)^2 - \mu a^4 -
\beta^2 \right]\psi (a) = 0.\end{equation}

\noindent We further define rescaled variables.

\begin{equation}\tilde a \equiv \alpha a\end{equation}

\begin{equation} \tilde \beta \equiv \beta / \alpha\end{equation}

\begin{equation}\tilde \gamma \equiv \alpha \gamma\end{equation}

\begin{equation}\tilde \mu \equiv \mu /\alpha^6\end{equation}

The Wheeler-DeWitt equation now becomes
\begin{equation} \left [- {\tilde a}^{-p}
{\partial\over \partial \tilde a}a^p{\partial \over \partial \tilde a} +
2U(\tilde a) \right]\psi (\tilde a) = 0.\end{equation}

\begin{equation}2U(\tilde a) \equiv (\tilde a-{\tilde \gamma})^2 -
\tilde \mu {\tilde a}^4 - {\tilde \beta}^2. \end{equation}

\begin{flushleft}{\large \bf III. MODEL WITH $k > 0$, $\rho=\rho_{{\rm rad}} $}
\end{flushleft}
\setcounter{chapter}{3}
\setcounter{equation}{0}

   In this section, we would like to study the Wheeler-DeWitt equation for
a closed FRW metric with just radiation as matter. There are several
reasons for studying this simplified case.  First, it is solvable in a
closed form.   Second, while  solving this model, we will encounter
general properties of quantizing a bound state system, i.e., existence
of a quantization condition and no need for an arbitrary boundary
condition for the wave function.

     The dimensionless parameters (2.16-2.19) are now $ \tilde \gamma =
\tilde \mu = 0$, $\tilde \beta \neq 0$, and the  Wheeler-DeWitt equation
(2.20,2.21)  is

\begin{equation}\left [ {d^2\over d{\tilde a}^2} + {p\over \tilde a}{d\over
d\tilde a}
- 2U(\tilde a)\right]\psi (\tilde a) = 0,\end{equation}

\noindent with   (Fig. 1)

\begin{equation}2U(\tilde a) \equiv {\tilde a}^2  - {\tilde \beta}^2.
\end{equation}

\noindent The dominant behavior of $\psi$ for $\tilde a \rightarrow \infty$
is $e^{-{\tilde a}^2/2}$.  Factoring  out the exponential behavior,
$\psi (\tilde a) \equiv g(\tilde a)e^{-{\tilde a}^2/2}$, $g(\tilde a)$
satisfies

\begin{equation}{d^2g\over d{\tilde a}^2} + ({p\over \tilde a} -2 \tilde a)
{dg\over d\tilde a}  +  ({\tilde \beta}^2 -1 -p )g = 0.\end{equation}

\noindent Power series method reveals that for $p\neq 0$ the function
$g(\tilde a)$ is an even function.  Therefore, in terms of $x \equiv
{\tilde a}^2$, $g(x)$ satisfies

\begin{equation}x{d^2g\over dx^2} + ( {1+p\over 2} -x){dg\over dx}  +
 ({{\tilde \beta}^2 -1 -p\over 4} )g = 0.\end{equation}

\noindent We first seek solution for the Laplacian factor ordering
(e.g. $p=1$)  [9, 20, 21]. For the wave function $\psi$ to remain finite
for $x \rightarrow \infty$, the parameter $\tilde \beta$  must
satisfy a quantization condition

\begin{equation}{\tilde \beta}^2 = 4n +2 \ \ n = 0,1,2...\end{equation}

\noindent The ``eigenfunctions'' are the familiar Laguerre polynomials  [22]

\begin{equation}g(x) = L^0_{n = {{\tilde \beta}^2 -2\over 4}}(x).\end{equation}

\noindent And finally, the wave functions are
\begin{equation}\psi_n(\tilde a) = e^{-{\tilde a}^2/2}L^0_{n =
{{\tilde \beta}^2 -2\over 4}}({\tilde a}^2).\end{equation}

\noindent In terms of radiation energy density, the quantization condition is

\begin{equation}\beta^2 \equiv ({3\pi\over 2Gk})^2{8\pi G\over 3k}
\rho_r a_o^4 = 4n+2.\end{equation}

\noindent The  eigenfunctions $\psi_n$, corresponding to first few
$n$ states are plotted in Fig. 2.

Before leaving this model, we would  like to investigate the  consequences
of  quantizing  the same system for  another operator ordering ($p=2$).
For $\rho = \rho_{{\rm rad}}$, the $p=2$ ordering is also exactly solvable.
The comparison of the two sets of solutions will lend credence to the
conventional wisdom   that  operator ordering may be dictated by convenience.

For $p=2$, the wave equation (3.1,3.2) can be rewritten in a form that
resembles a one-dimensional wave equation with a potential.
$g(\tilde a)$  defined as  $\psi (\tilde a) = g(\tilde a)/\tilde a$ satisfies

\begin{equation}\left [ {d^2\over d{\tilde a}^2} - {\tilde a}^2  +
{\tilde \beta}^2 \right]g(\tilde a) = 0,\end{equation}

\noindent which is the familiar Schrodinger equation for a harmonic
oscillator. The only difference is that among the complete set of
Hermite polynomials,
which makes up the solution to harmonic oscillator problem,  only
the odd Hermite polynomials  are acceptable in this case.  This is
necessary in order for the wave function
$\psi(\tilde a) \equiv {g(\tilde a)\over \tilde a}$
to remain finite at $\tilde a =0$. The familiar quantization condition for
harmonic oscillator gives

\begin{equation}{\tilde \beta}^2 = 2({\rm odd}) +1 = 4n + 3, \ \ n = 0,1,2
...\end{equation}

\noindent Comparison with the result for the Laplacian ordering (3.5) shows
that the eigenvalue spectrum is essentially unchanged.  It can also be
shown that for an arbitrary $p \geq 1$ operator ordering, the
quantization condition is

\begin{equation}{\tilde \beta}^2 = 4n + 1 + p, \ \ n = 0,1,2 ...\end{equation}

\begin{flushleft}{\large \bf IV. MODEL WITH $k > 0$, $\rho=\rho_{{\rm rad}}
+ \rho_{{\rm dust}}$}
\end{flushleft}
\setcounter{chapter}{4}
\setcounter{equation}{0}

In this section, we would like to study the Wheeler-DeWitt equation for a
more realistic case of FRW metric with a combination of radiation and
pressureless dust.

 The dimensionless parameters (2.16-2.19) are now $ \tilde \gamma \neq 0$,
$\tilde \beta \neq 0$, and $\tilde \mu = 0$. To keep the model as realistic
as possible, we will assume that
$\rho_{{\rm rad}}/\rho_{{\rm dust}} \approx  O(10^{-4})$ at $a_o\equiv 1$,
which implies that  $\tilde \beta \approx \tilde \gamma$.

The  Wheeler-DeWitt equation (2.20,2.21)  is  now

\begin{equation}\left [ {d^2\over d{\tilde a}^2} + {p\over \tilde a}
{d\over d\tilde a}
- 2U(\tilde a)\right]\psi (\tilde a) = 0,\end{equation}

\noindent with

\begin{equation}2U(\tilde a) \equiv (\tilde a-{\tilde \gamma})^2 -
{\tilde \beta}^2.\end{equation}

 A schematic graph of the ``potential'' $U(\tilde a)$ is given in Fig. 3.
The solution to this differential equation doesn't seem to be
analytically tractable for the Laplacian operator ordering ($p=1)$. We
will therefore resort to solving the equation for another ordering
(i.e., $p=2$) where the problem is analytically tractable.

Again for $p=2$, the wave equation (4.1,4.2) can be rewritten in a
form that resembles a one-dimensional wave equation with a potential.
$g(\tilde a)$ defined as $\psi (\tilde a) = g(\tilde a)/\tilde a$ satisfies

\begin{equation}\left [ {d^2\over d{\tilde a}^2} -
(\tilde a-{\tilde \gamma})^2 + {\tilde \beta}^2 \right]g(\tilde a) = 0,
\end{equation}

\noindent which resembles a simple harmonic oscillator with a shift
$\Delta x =  - \tilde \gamma$ in the potential.  The key difference is
that here the range of coordinate is $\tilde a \in [0, \infty]$ whereas
for the  harmonic oscillator the range is $[-\infty, \infty]$.

A solution to (4.3) which vanishes for very large positive values of
$\tilde a $ is

\begin{equation}g(\tilde a) = D_{ {\tilde \beta}^2 - {1\over 2}}
(\tilde a - \tilde \gamma),\end{equation}

\noindent where $D_{\nu}(z)$ is the parabolic cylinder function [23].

The quantization condition is obtained by requiring  that

\begin{equation}g(\tilde a = 0) =  D_{ {\tilde \beta}^2 -
{1\over 2}}(- \tilde \gamma) = 0. \end{equation}

\noindent This is again necessary in order for  the wave function
$\psi (\tilde a) \equiv g(\tilde a)/\tilde a$ to remain finite at
$\tilde a =0$.  Disappointingly, it is difficult to calculate the
parameters $\tilde \beta$, and $\tilde \gamma$ for which (4.5) is satisfied.
We are left to resort to WKB approximation to find the approximate
``eigenvalues''.  The appropriate WKB phase integral relevant for (4.3) is

\begin{equation}J \equiv \int_0^{\tilde \beta + \tilde \gamma}
(-2U(\tilde a))^{1/2}d\tilde a = \pi (n + {3/4}) \ \ n=0,1,2.. \end{equation}

\noindent The usual $1/2$ is replaced by $3/4$ because  $\tilde a = 0$ is
not a usual classical turning point but an end of the coordinate, and  also
because we must demand that $g({\tilde a}=0) = 0$ [24]. For the limiting
case of $\tilde \beta = \tilde \gamma$, which is the statement that
$\rho_{{\rm rad}} = 0$ and $\rho=\rho_{\rm dust}$, the phase integral can
be readily evaluated to give $J = {\pi\over 2}{\tilde \gamma}^2$.

In terms of $\rho = \rho_{\rm dust}$, the quantization condition is

\begin{equation}({3\pi\over 2Gk})({8\pi G\over 6k}\rho_d a_o^3)^2 =
2(n + 3/4).\end{equation}

\noindent For our Universe, $\tilde \beta \neq \tilde \gamma$ but

\begin{equation}{ {\tilde \beta }^2 - {\tilde \gamma}^2\over {\tilde\gamma}^2}
 \approx {\rho_r\over \rho_d}{2k\over H_0^2 +k} \sim O(10^{-4}).\end{equation}

\noindent In the Appendix , the phase integral for
$\tilde \beta \neq \tilde \gamma$ case is evaluated  to give
$J \approx {\pi\over 2}{\tilde \beta}^2$. In terms of $\rho_{\rm rad}$ and
$\rho_{\rm dust}$, the quantization condition is

\begin{equation} ({3\pi\over 2Gk})\left( {8\pi G\over 3k}\rho_r a_o^4 +
({8\pi G\over 6k}\rho_d a_o^3)^2\right) = 2(n + 3/4).\end{equation}

\noindent If we make a rough estimate that $k \sim H_o^2$, then the
quantum number of our Universe is

\begin{equation}N \sim (G k)^{-1} \sim \left ({{\rm three  \ radius}\over
{\rm Planck  \ length}}\right )^2 \sim \left ({H_o^{-1}\over
{\rm Planck \ length}}\right )^2 \sim 10^{122}.\end{equation}

A comment is in order.  It is obvious that the entire analysis was
facilitated by the use of $p=2$ operator ordering.  It is also easy to
convince oneself that even if we could solve for the $p=1$ operator
ordering  the two quantization results should  not differ by much.  The
argument goes as follows.  Because the operator ordering parameter ($p$)
appears with  ${\tilde a}^{-1}$ in the Wheeler-DeWitt equation (4.1), a
wrong value  for $p$ should affect the small range of $\tilde a$ the most.
But for a universe with a combination of radiation and dust, small value
for the scale factor corresponds to  the radiation dominated era.  And
since the two exact  quantization conditions  for $p=1,2$  orderings
with $\rho = \rho_{\rm rad}$ are essentially the same (3.5,3.10), we can
also safely conclude that  the quantization conditions for $p=1,2$ orderings
should not differ by  much even for the present case.

Interestingly, the unusual largeness of the quantum number of our Universe
can be easily explained within the framework of inflationary universe
models.  This will be addressed in the next section.

\begin{flushleft}{\large \bf V. MODEL WITH $k > 0$,
$\rho=\rho_{\rm rad} + \rho_{\rm dust}$, and $\Lambda \neq 0$}
\end{flushleft}
\setcounter{chapter}{5}
\setcounter{equation}{0}

In this section, we would like to study the effects of including a
cosmological constant to a FRW universe which in its late stage of
evolution becomes radiation and then matter dominated. A straightforward
method, at least formally,  would be to include a spatially homogeneous
scalar field $\phi$ with some self coupling $V(\phi)$, whose non-zero
expectation value would play the role of a cosmological constant [25].
Moreover, as a reheating mechanism, there should also be some coupling
between the scalar field and the radiation field, so that after the
$e^{60}$ expansion the scalar field could transfer its vacuum energy to the
radiation field.  Needless to say this is a very tall order.   Here we make
a loose-shoe approximation by putting in by hand a cosmological constant
$\Lambda_{\rm GUT}$.  And to simulate  the reheating mechanism, we
set $\Lambda_{\rm GUT} = 8\pi G\rho_{\rm rad}$ at some high redshift
(e.g., $z = z_{\rm GUT}$) and then set $\Lambda_{\rm GUT} = 0$ for
$z \leq z_{\rm GUT}$.

The Wheeler-DeWitt equation is now (2.16-2.21)

\begin{equation}\left [ {d^2\over d{\tilde a}^2} + {p\over \tilde a}
{d\over d\tilde a} - 2U(\tilde a) \right]\psi (\tilde a) = 0,\end{equation}

\noindent with

\begin{equation} 2U(\tilde a) = \left \{ \begin{array}{ll}
{\tilde a}^2 - {\tilde \mu}{\tilde a}^4 \ \ \  &{\rm z\geq z_{GUT}} \\
(\tilde a-{\tilde \gamma})^2 - {\tilde \beta}^2 & {\rm z \leq z_{GUT}}
\end{array}\right. \end{equation}

A schematic plot of $U(\tilde a)$ is given is Fig. 4.  From the figure,
we can readily conclude that it is again a bound state problem. Moreover,
we can immediately infer an essential  difference between including and
not including the reheating phase.  If there were no reheating but a
constant $\Lambda \neq 0$ for all time, then the potential $U(\tilde a)$
would describe a non-bound state problem Fig. 5.  Moreover,
$\Lambda \rightarrow 0$   has a unique role in the parameter space.
Intuitively, if $\Lambda_{\rm GUT}$ settles to some small but a finite
value $\Lambda_{\rm final}$, then the metric can re-enter into a
de-Sitter phase, albeit the two de-Sitter phases may be separated by a
classically forbidden region. Mathematically, this is because
$\Lambda$ appears in $U(\tilde a)$ with the ${\tilde a}^4$ factor, which
eventually comes to dominate $U(\tilde a)$. In the Wheeler-DeWitt formalism,
this would correspond to a tunneling model, which is analogous to a quantum
mechanical description of Alpha decay of a heavy nucleus. A schematic plot of
classically allowed regions in $U(\tilde a)$ for this case is given in Fig. 6.

If we assume that  we live in this classical allowed region with a small
$\Lambda_{\rm final} \sim O(1) H_o^2$, then it is very interesting to
estimate the tunneling probability. The Euclidean action for tunneling
region is  [26]

\begin{equation}S_E \approx - (G\Lambda_{\rm final} )^{-1}  \approx -
G^{-1}H_o^{-2} \approx -122.\end{equation}

\noindent And the tunneling probability is [27-30]

\begin{equation}P_T \propto exp(-\vert S_E\vert) \sim exp(-122).\end{equation}

Now let us return to the case where $\Lambda_{\rm GUT} \neq 0$ only at
large redshifts (e.g. $z \geq z_{\rm GUT}$) and $\Lambda_{\rm final} = 0$.
We are interested in its effect on the quantization condition (4.9).  It is
easy to trace the large $n \sim 10^{122}$ back to the factor $Gk$.  In
light of this fact, having a cosmological constant in our early universe
can be  interpreted as the mechanism which ``generated'' this large quantum
number. This is because one of the triumph of an inflationary universe model
is its ability to naturally explain the flatness of the observed Universe.

Finally, let us investigate the nature of $\psi$ at early Universe.
For $z \ll z_{\rm GUT}$, the wave function should be that of
$\rho = \rho_{\rm rad} + \rho_{\rm dust}$  model (4.4).  For $z \ll
z_{\rm GUT}$ Wheeler-DeWitt equation for the Laplacian ordering is (5.1,5.2)

\begin{equation}\left [ {d^2\over d{\tilde a}^2} + {1\over \tilde a}
{d\over d\tilde a} + {\tilde a}^2 -\tilde \mu {\tilde a}^4 \right]\psi
(\tilde a) = 0.\end{equation}

\noindent Moreover, near $\tilde a \approx 0$ the cosmological constant
term is negligible compare to the curvature term.  For this region, the
differential equation has a single independent solution

\begin{equation}\psi (\tilde a) = I_0({\tilde a}^2/2).\end{equation}

\noindent $I_{\nu}(x)$ is a modified Bessel functions [22].  Its power
series expansion  is

\begin{equation}I_{\nu}(x) = \sum_{s=0}^{\infty}{1\over s!(s+\nu)!}\left
({x\over 2}\right )^{2s+\nu}.\end{equation}

\noindent Interestingly, $\psi (\tilde a = 0) \neq 0$ because $I_0(0) = 1$.

There are also inflationary universe models where $\rho_{rad} \neq 0$ before
the inflationary phase  [31,32].   In such a case, the Wheeler-DeWitt
equation for $z \geq z_{\rm GUT}$ is (2.16-2.21)

\begin{equation}\left [ {d^2\over d{\tilde a}^2} + {1\over \tilde a}
{d\over d\tilde a} +
{\tilde a}^2 -\tilde \mu {\tilde a}^4 - {\tilde \beta}^2\right]\psi
(\tilde a) = 0.\end{equation}

\noindent Again near $\tilde a \approx  0$ the cosmological constant term is
negligible.  The resulting differential equation has a single independent
solution (3.7)

\begin{equation}\psi_n(\tilde a) = e^{-{\tilde a}^2/2}L^0_{n =
{{\tilde \beta}^2 -2\over 4}}({\tilde a}^2).\end{equation}

\noindent And because $L_n(0) = 1$ the wave function $\psi_n({\tilde a}=0)
\neq 0$.

\begin{flushleft}{\large \bf VI. DISCUSSION}
\end{flushleft}
\setcounter{chapter}{6}
\setcounter{equation}{0}

In this section we would like to briefly discuss an interesting
generalization.  Consider  a ($k > 0$) mini-superspace model, which ends
in $\rho = \rho_{\rm rad} + \rho_{\rm dust}$, but now  the corresponding
Wheeler-DeWitt Hamiltonian is invariant under some symmetry transformation
by having more than one gravitational degree of freedom, e.g., homogeneous
but anisotropic Kasner metric  [33-35]. In general, symmetry is a sign of
degenerate in eigenvalues. In such a case we would have an interesting
consequence.  First, the problem is still a bound state, so there would be
no need for an arbitrary boundary condition when solving for the wave
functions.  But from general quantum mechanical considerations, eigen-wave
functions  for a bound state is guaranteed to be real only if there isn't
any degeneracy. So we have a combination of Hartle-Hawking's no boundary
condition with Vilenkin's complex wave functions with it non-zero flux,
$J = {1\over 2}(\phi \nabla \psi^* - \psi^*\nabla\psi)$. The situation
would be analogous to $n^2$ degeneracy for each $E_n$ level in a hydrogen
atom.

\begin{flushleft}{\large \bf VII. CONCLUSION}
\end{flushleft}
\setcounter{chapter}{7}
\setcounter{equation}{0}

In this paper, we have studied the solutions to Wheeler-DeWitt equation for
$k > 0$ Friedmann Robertson Walker metric with various types of matter.
First, we have  shown that if the Universe ends in the  matter dominated era,
then the resulting Wheeler-DeWitt equation describes a bound state problem.
As solutions of a non-degenerate bound state system, the eigen-wave functions
are  real (Hartle-Hawking) and the usual issue associated with the
ambiguity in the boundary conditions for the wave functions is resolved.
Furthermore, as a bound state problem, there exists a quantization
condition that relates the curvature of the three space with the energy
density of the Universe. Incorporating a cosmological constant in the
early Universe (inflation) was given as a natural explanation for the
large quantum number our Universe, which resulted from the quantization
condition. Second, we have shown  that if there is a cosmological
constant $\Lambda > 0$  in our Universe that persists for all time, then
the resulting Wheeler-DeWitt equation  describes a non-bound state system,
regardless of the magnitude of the cosmological constant.  As a consequence,
the wave functions are in general complex (Vilenkin) and the initial
conditions for   wave functions are  free parameters not determined by the
formalism.

\begin{center}{\large \bf  APPENDIX}
\end{center}
\setcounter{chapter}{8}
\setcounter{equation}{0}

We would like to show that

\begin{eqnarray*} J \equiv \int_0^{{\tilde \beta} + {\tilde \gamma}} p dx
\approx {\pi\over 2}{\tilde \beta}^2\end{eqnarray*}

\noindent
with
\begin{eqnarray*} p = \left({\tilde \beta}^2 - ({\tilde \gamma - \tilde a})^2
\right)^{1/2}. \end{eqnarray*}
\noindent
For our Universe
\begin{eqnarray*} {{\tilde \beta}^2 -{\tilde \gamma}^2\over
{\tilde \gamma}^2} \approx {2k\over H_0^2 + k}{\rho_{\rm rad}\over
\rho_{\rm dust}} \sim O(10^{-4}).\end{eqnarray*}

\noindent Therefore,  defining $x = {\tilde a}/{\tilde \gamma}$, we have

\begin{eqnarray*}
J &=&  \int_0^{{\tilde \beta} + {\tilde \gamma}} \left({\tilde \beta}^2 -
({\tilde \gamma - \tilde a})^2\right )^{1/2}dx \\
&=& \int_{\tilde \gamma}^{\tilde \beta}{\tilde \gamma }^2\left(
({\tilde \beta\over \tilde \gamma })^2 - ({\tilde a\over \tilde \gamma} )^2
\right )^{1/2}
d({\tilde a\over \tilde \gamma}) \\
&\approx&  {\tilde \gamma }^2\int_{-1}^{1}\left (
({\tilde \beta\over \tilde \gamma})^2 - x^2 \right )^{1/2}dx \\
&\approx&  {\tilde \gamma }^2\int_{-1}^{1}\left[ (1 - x^2)^{1/2}  +
{1\over 2}{ ({\tilde \beta \over \tilde \gamma })^2 -
1 \over  (1 - x^2)^{1/2} }            \right ]dx \\
&\approx&  {\tilde \gamma}^2\left [ \pi /2 +
\pi /2\left (({\tilde \beta \over \tilde \gamma })^2 - 1 \right )\right] \\
&\approx&  {1\over 2} \pi {\tilde \beta}^2.\ \
\end{eqnarray*}

\begin{center}{\large \bf IX. ACKNOWLEDGMENTS}
\end{center}
\setcounter{chapter}{9}
\setcounter{equation}{0}
I would like to thank George Field, Christina Doyle, Eric Blackman, Min Yan
for making my extended stay at the Center for Astrophysics so pleasant.
\newpage

\begin{center}
{\large \bf FIGURE CAPTIONS}
\end{center}
\begin{enumerate}
\item Graph of $U(\tilde a)$ for $\rho = \rho_{\rm rad}$ and $k > 0$.
\item Graphs of  $\psi_n(\tilde a)$ for first few quantum states. Figures
2(a)-2(d) correspond to $n=[0,1,2,20]$, respectively. The normalization is
arbitrary.
\item Graph of $U(\tilde a)$ for $\rho = \rho_{\rm rad} + \rho_{\rm dust}$
and $k > 0$.
\item Graph of $U(\tilde a)$ for $\rho = \rho_{\rm rad} + \rho_{\rm dust}$,
$k > 0$, and $\Lambda_{\rm GUT} \neq 0$ for $z \geq z_{\rm GUT}$.
\item Graph of $U(\tilde a)$ for $k > 0$ and $\Lambda \neq 0$ for all time.
\item Graph of $U(\tilde a)$ for $\rho = \rho_{\rm rad} + \rho_{\rm dust}$,
$k > 0$,  $\Lambda_{\rm GUT}$, and $\Lambda_{\rm final}$.
$\Lambda_{\rm GUT}$  is responsible for inflation and $\Lambda_{\rm final}$
is some final value the cosmological constant has settled into in a universe.

\end{enumerate}

\newpage
\begin{center}
{\large \bf REFERENCES}
\end{center}
\begin{enumerate}
\item P. A. M Dirac, {\it Lectures on Quantum Mechanics}
(Belfer Graduate School of Science Monographs Number Two, Yeshiva University,
New York, 1964).
\item R. M. Wald, {\it General Relativity} (Univ. of Chicago Press, Chicago,
1984).
\item R. Arnowitt, S. Deser, and C. W. Miser,  in {\it Gravitation: An
Introduction to Current Research}  (Wiley, New York, 1963).
\item B. S. DeWitt, Phys. Rev. {\bf 160}, 1113 (1967).
\item J. A. Wheeler, in {\it Battelle Rencontres}, edited by C. DeWitt and
J. A. Wheeler (Benjamin, New York, 1968).
\item C. W. Misner, in {\it Magic Without Magic: John Archibald Wheeler,}
edited by J.R. Klauder (Freeman, San Francisco, 1972).
\item J. J. Halliwell, Phys. Rev. {\bf D 38}, 2468 (1988).
\item C. W. Misner, in {\it Relativity,} edited by M Carmeli, S. I. Fickler,
and L. Witten (Plenum, New York, 1970).
\item S. W. Hawking and D. N. Page, Nucl. Phys. {\bf B 264}, 185 (1986).
\item W. F. Blyth and C. J. Isham, Phys. Rev. {\bf D 11}, 768 (1975).
\item T. Christodoulakis and J. Zanelli, Phys. Lett. {\bf 102 A}, 227 (1984).
\item G. Esposito and G. Platania, Class. Quantum Grav. {\bf 5}, 937 (1988).
\item G. W. Gibbons and L. P. Grishchuk, Nucl. Phys. {\bf B313}, 736 (1988).
\item J. B. Hartle and S. W. Hawking, Phys. Rev. {\bf D 28}, 2960 (1983).
\item S. W. Hawking, Nucl. Phys. {\bf B 239}, 257 (1984).
\item A. Vilenkin, Phys. Rev. {\bf D 33}, 3560 (1986).
\item A. Vilenkin, Phys. Rev. {\bf D 37}, 888 (1986).
\item E. Alvarez, Rev. Mod. Phys. {\bf 61}, 561 (1989).
\item C. W. Misner, K. S. Thorne, and J. A. Wheeler, {\it Gravitation}
(Freeman, San Francisco, 1973).
\item G. A. Ringwood, J. Phys. A: Math. Gen., {\bf 9}, 1253 (1976).
\item K. S. Cheng, J. Math. Phys., {\bf 13}, 1723 (1972).
\item G. Arfken, {\it Math. Methods for Physicists} (Academic, New York, 1970).
\item I. S. Gradshteyn and I. M Ryzhik, {\it Table of Integral, Series, and
Products} (Academic, New York, 1965).
\item E. Merzbacher, {\it Quantum Mechanics} (Wiley, New York, 1970).
\item A. Guth, Phys. Rev. {\bf D 23}, 347 (1981).
\item E. W. Kolb and M. S. Turner, {\it The Early Universe}
(Addison-Wesley, New York, 1990).
\item A. D. Linde, JETP {\bf 60}, 211 (1984).
\item A. Vilenkin, Phys. Rev. {\bf D 30}, 549 (1984).
\item V. A. Rubakov, Phys. Lett. {\bf 148 B}, 280 (1984).
\item Ya. B. Zel'dovich and A. A. Starobinskii, Sov. Astron. Lett.
{\bf 10}, 135 (1984).
\item A. Albrecht and R. Brandenberger, Phys. Rev. {\bf D 31}, 1225 (1985).
\item J. H. kung and R. Brandenberger, Phys. Rev. {\bf D 40}, 2532 (1989).
\item E. Kasner, Am. J. Math. {\bf 43}, 217 (1921).
\item E. Schucking and O. Heckmann, "World models," in {\it Onzieme Conseil
de Physique Solvay}, Editions Stoops, Brussels (1958).
\item C. W. Misner, Phys. Rev. Lett. {\bf 22}, 1071 (1969).

\end{enumerate}

\end{document}